\documentclass[longbibliography,twocolumn,showkeys,superscriptaddress,secnumarabic,amssymb, nobibnotes, aps, prd]{revtex4-1}
\usepackage{upgreek}
\usepackage{graphicx}

\begin{document}

\title{Identifying signature Zernike modes for efficient light delivery through brain tissue}%

\author{Sharmila S. San\'e}%
\affiliation{John Curtin School of Medical Research, Australian National University, 0200 ACT, Australia
}
\author{Julian M.C. Choy}%
\affiliation{John Curtin School of Medical Research, Australian National University, 0200 ACT, Australia
}
\author{Woei Ming Lee}%
\affiliation{Research School of Engineering, Australian National University, 0200 ACT, Australia
}
\author{Christian Stricker}%
\affiliation{John Curtin School of Medical Research, Australian National University, 0200 ACT, Australia
}
\affiliation{Medical School, Australian National University, Canberra, 0200 ACT, Australia
}
\author{Hans A. Bachor}%
\affiliation{Research School of Physics and Engineering, Australian National University, 0200 ACT, Australia
}

\author{Vincent R. Daria}%
\email{vincent.daria@anu.edu.au}
\affiliation{John Curtin School of Medical Research, Australian National University, 0200 ACT, Australia
}
\date\today%

\begin{abstract}
Recent progress in neuroscience to image and investigate brain function has been made possible by impressive developments in optogenetic and opto-molecular tools. Such research requires advances in optical techniques for the delivery of light through brain tissue with high spatial resolution. The tissue causes distortions of the wavefront of the incoming light which broadens the focus, thereby reducing the intensity and resolution especially in techniques requiring focal illumination. Adaptive wavefront correction has been demonstrated to compensate for these distortions. However, in many situations iterative derivation of the corrective wavefront introduces time constraints that limit its usefulness when used to probe living cells. Here we demonstrate a direct and fast technique by working with a small set of Zernike modes and demonstrate that corrections derived \textit{a priori} can lead to significant improvement of the focus. We verify this idea by the electrical response of whole-cell patched neurons following two-photon photolysis of caged neurotransmitters along its dendrites. In particular, we find that the organization of the neuropile in the cortical region of rat brain slices provide sufficient \textit{a priori} information to preselect an effective wavefront correction.
\end{abstract}


\maketitle

\section{Introduction} 
Understanding the fundamental processes of neurons that are embedded deep within brain tissue remains a demanding task, as existing optical microscopy techniques do not perform well  beyond the superficial layers. While two-photon microscopy uses a less absorptive near-infrared (NIR) laser, refractive index inhomogeneity and scattering are still important factors that distort the point spread function (PSF) of the incident coherent light. Aside from affecting the resolution, the broadening of the PSF drastically reduces the probability of non-linear two-photon absorption and consequently decreases fluorescence emission. While the signal can be provisionally corrected by increasing the intensity of the incident light, this can result in photo-toxicity and heating of the tissue \cite{gautam2015improved}. Such unwanted effects can affect physiological activity and are hence disruptive when probing fundamental cellular processes. Fluorescence imaging and photostimulation (e.g. photolysis of chemically caged compounds and optogenetics) will all suffer the same predicament.  Hence, the key to achieve minimally invasive imaging and probing of cellular processes is to build-up the capacity to rectify distortions of the incident light field, which optimizes the ability to achieve a near diffraction-limited focus in deep regions of the brain tissue.\\

Although healthy brain tissues are made up of a heterogeneous distribution of neuronal types, dendrites and axons, these tissues are generally built up of a specific organization of cells and subcellular compartments. For example in the neocortex, a diverse spread of neurons organize into several layers forming the neuronal circuitry for information processing in the brain \cite{bullmore2012economy}. When viewed at a certain scale, the ordered arrangement of cells in tissue presents a predicable framework for mapping certain regions and targeting specific cells. \\

Earlier reports of optical imaging through optically thick brain tissues has paved the way to our investigations of neurons in their natural environment. When imaging optically thick brain tissues, light encounters multiple scattering to deviate from its original path. The scattered optical paths could be treated as almost deterministic and, when measured accordingly, it would be possible to build a transmission matrix \cite{yaqoob2008optical} or phase map \cite{schwertner2004characterizing} to correct for the light distortion in that particular instance.  As such, recent techniques into the measurement and reversal of optical distortion in tissues \cite{booth2007adaptive} based on photo conjugation \cite{yaqoob2008optical}, turbid layer conjugation \cite{park2015high}, sensorless imaging \cite{ji2010adaptive,wang2014rapid}, artificial guide star \cite{tao2011fluor} and wavefront sensors \cite{tao2011adaptive} have shown promising results. These techniques correct for the light distortion by treating sections of the tissue as distinct transmission matrices and hence require pre-measurement of a corrective wavefront for the incoming light prior to obtaining a wavefront corrected high-resolution image. In general, such active measurement works well for biological samples that do not deteriorate with time. However, as for many experiments involving living cells, when the window of time for the right experimental conditions is limited, an \textit{a priori} identification of an appropriate wavefront correction is necessary.   \\

In this work, we identify the signature phase correcting maps for particular regions in optically thick brain tissue. This allows us to pre-correct for light distortion without any wavefront sensing  \cite{schwertner2004characterizing} by using predictable Zernike modes measured \textit{a priori}.  While light experiences multiple scattering in tissues, differential interference contrast (DIC) imaging of these tissues relay that they still converge to form perceptible images giving us information of the cellular network.  As cells within tissue can be distinguished by changes in refractive index, the cellular network can form visible and highly organized structures of varying refractive indices that manifest as optical aberrations for the incoming light. We can therefore correct for these aberrations using Zernike modes. \\

For the experimental application, we require rapid optimization of the laser focus through freshly prepared parasagittal brain slices from a rat for \textit{in vitro} opto-electrophysiological experiments. Before performing experiments with living cells, we first used fixed tissue samples to find signature Zernike modes that persistently optimize the focus at different locations within a selected brain region. It was apparent from the iterative procedure that a small subset of the modes were responsible for optimizing the focus. Hence we hypothesized that such a set of Zernike modes may be associated with specific brain regions and could hence be used for wavefront correction in that region for different animals of the same age and species. This is demonstrated here by improving the efficiency of two-photon photolysis along dendrites of neurons embedded within brain slices. Two-photon photolysis releases chemically caged neurotransmitters (glutamate) near dendritic spines, thus emulating synaptic inputs to the neuron \cite{callaway1993photostimulation, denk1994two}. We show that there is an optimum uncaging response of caged glutamate on a select set of Zernike modes encoded on the excitation light. Using just these few pre-determined Zernike modes associated with specific brain regions allows the bulk of the wavefront correction to be made with minimal optimization, which is advantageous  in time-critical experiments where the lengthy search for the ideal wavefront correction is not feasible.\\

\begin{figure*}[!]
\includegraphics[width=12cm]{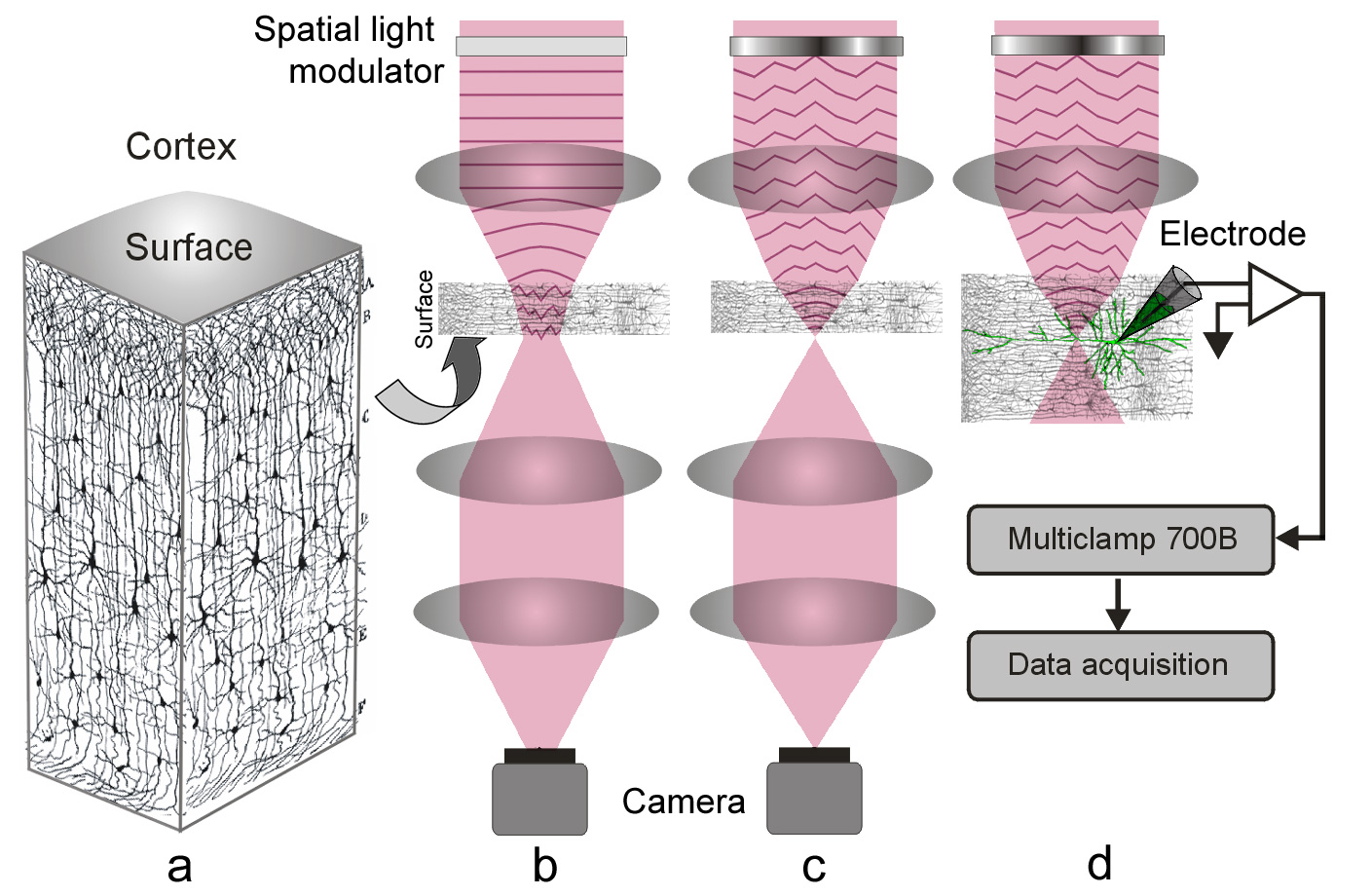}
 \caption{\label{figure1} Schematic of the experiment. (a) A 3D section of a cortical brain slice (adapted and modified from Cajal, 1909 \cite{cajal1995histology}), showing organization of the neuropile. (b) Uncorrected light is scattered as it enters the brain tissue thus broadening the focus. (c) The spatial light modulator (SLM) is encoded with a corrective wavefront to compensate the aberrations introduced by the tissue, resulting in a sharp focus. In (b) and (c) the focus at the back of the sample is imaged by a camera. (d) The corrected focus within the sample allows for optimal photostimulation with the neuronal response recorded by the electrode. }
 \end{figure*}

\section{Methodology}
\subsection{\textit{A priori} identification of Zernike modes}

After calibrating the system with optical materials of known optical aberration (see appendix A), we proceed to optimize the laser focus through fixed brain tissue.  Figure \ref{figure1} shows a schematic of this experiment.  Figure \ref{figure1}a shows a graphical illustration of a cortical slice adapted from Cajal \cite{cajal1995histology}. We fixed 100 $\upmu$m and 300 $\upmu$m thick parasagittal brain slices from 15 to 19 day old Wistar rats (see Appendix B for brain slice preparation).  The slices were placed in between two type-0 coverslips and observed under a custom-built microscope described in Appendix C. The fixed brain slices were used for prior determination of the Zernike mode correction schematically described in figure \ref{figure1}b, which shows an uncorrected beam propagating through the tissue and figure \ref{figure1}c showing a patterned beam wavefront corrected via a spatial light modulator (SLM). \\

\begin{figure}
\includegraphics[width=8.5cm]{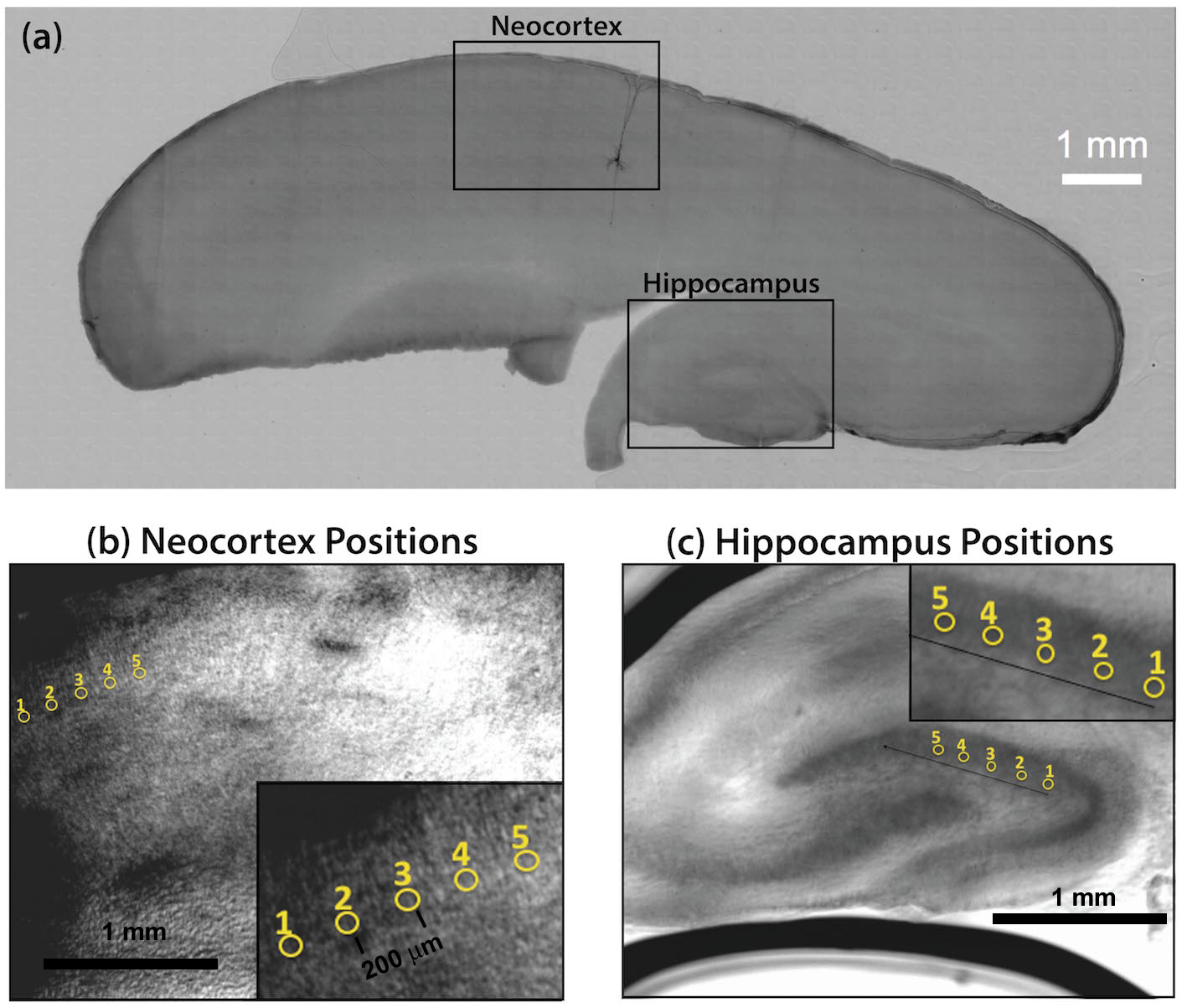}
 \caption{\label{figure2} (a) Parasagittal rat brain slice indicating the neocortex and hippocampus. A single biocytin-filled Layer V pyramidal neuron shows the typical orientation of similar type of neurons in the neocortex.  (b) 5 positions within the neocortex  and (c) 5 positions within the hippocampus from where optimization data was obtained.}
\end{figure}

\begin{figure}
\includegraphics[width=8.5cm]{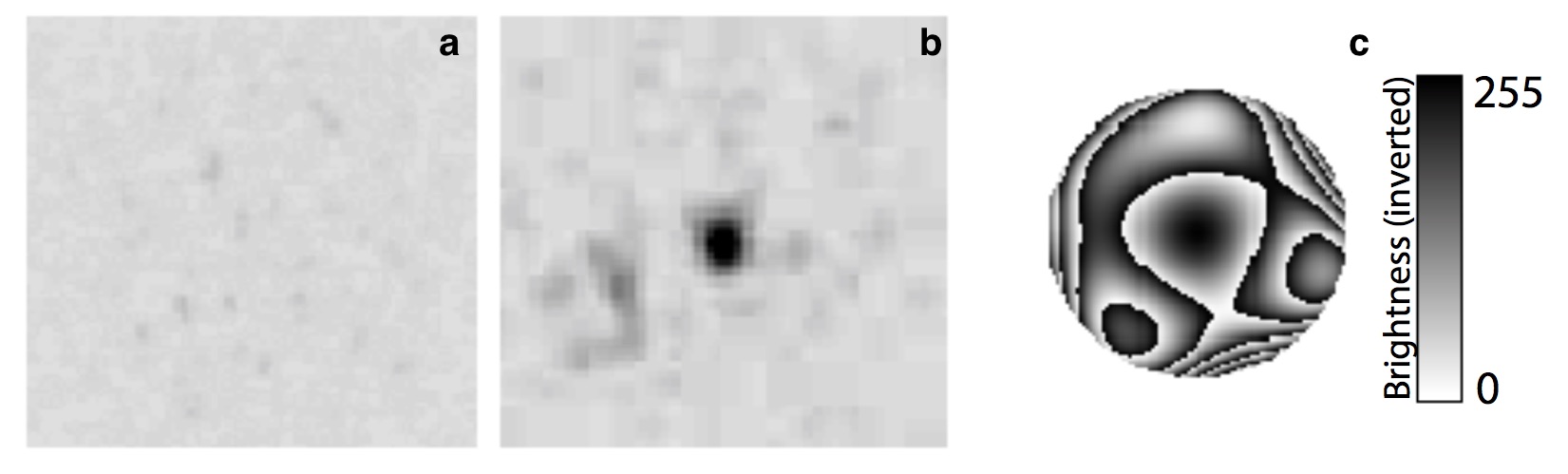}
 \caption{\label{figure3} A representative local correction in the neocortex at 100 $\upmu$m depth. (a) A broadened focus imaged at the bottom of the sample. (b)  Restored focus after encoding a corrective wavefront on the laser. (c) Phase pattern corresponding to the corrective wavefront showing a linear combination of Zernike modes at the resolution displayed on the SLM. Color bar shows black as the highest intensity for (a) and (b), and 2$\uppi$ for (c). }
\end{figure}

To find the appropriate corrective wavefront, an iterative algorithm was applied to maximize the beam intensity through the slice of brain tissue. We find corrective wavefronts on two key cortical regions in the brain slice, namely the neocortex and the hippocampus. These regions are frequently used in \textit{in vitro} studies of brain function. Sets of locally optimized phase corrective wavefronts in 5 separate positions at  around 200$\upmu$m apart were recorded (see figure \ref{figure2}). The metric used to retrieve the aberration correction was taken from the quality of the beam focus positioned at the bottom of the tissue sample.  The focus was imaged onto a camera and the quality of the beam was maintained by imposing a digital pinhole on the image of the focus.  As the incident laser propagates through the sample, the aberrations reduce the intensity of the focus. We iterate a finite number of Zernike modes (Noll Zernike terms, NZT = 1 to 15) over phase multipliers (or coefficients), which change the wavefront of the incident light and influence the intensity of the laser spot. Our `hill-climbing' algorithm incrementally alters the coefficient of each mode in an iterative loop, incorporating only the modes which increase the intensity of the focus. This correction is achieved by encoding the phase pattern onto the SLM, and actively tracking the net intensity within the digital pinhole after each change. The algorithm converges to a phase pattern producing a focal spot with high total intensity (see figure \ref{figure3}), and the coefficients of the Zernike modes corresponding to the correction are recorded. This method allows us to select the optimal configuration Zernike modes without resorting to pixel-based and arbitrary phase manipulation \cite{ji2010adaptive}.  \\

\subsection{Photolysis of caged neurotransmitters}
 
Figure \ref{figure1}d shows the schematic of how we improved the efficiency of uncaging by correcting the wavefront using the signature Zernike modes derived in the previous section.  A glass electrode was patched onto a cortical layer V pyramidal neuron embedded 50-100 $\upmu$m deep within the parasagittal brain slice (300 $\upmu$m thick).  The electrode contained an internal cellular solute as well as Alexa Fluor dye to visualize the dendritic tree (see Appendix B for details of the \textit{in vitro} experiment).  After patching, the dye was allowed to diffuse into the neuron for 20-30 min before imaging the neuron using a custom-built two-photon microscope described in Appendix C. To identify sites for photolysis (or uncaging) of caged neurotransmitters, we visualized the dendritic tree by rendering a 3D image of the neuron. From the 3D image, we identified sites for 2P glutamate uncaging along the dendritic tree and positioned a single uncaging spot via holographic encoding as described in our previous work \cite{go2013four}. MNI-caged glutamate (20 mM; Tocris Bioscience) was locally perfused through a micro-pipette introduced at a constant pressure (1.0 kPa). The laser pulse (2 ms) was controlled via a Uniblitz VS25 shutter (Vincent Associates). Two-photon imaging was performed at 800 nm wavelength with an average power set to 7-10 mW while uncaging was done at 720 nm wavelength with average powers of 20-25 mW measured at the focus. We worked down to depths of 50-150 $\upmu$m in the 300 $\upmu$m thick acute living brain slices. For imaging, the glutamate-puffing micropipette was kept out of the imaging field of view and no positive pressure was applied to the pipette to avoid premature uncaging. \\

The glass electrode used to fill the cell with Alexa dye was also used as the recording electrode. Recording of postsynaptic potentials and action potentials was done in current clamp mode using a MultiClamp 700B amplifier (Molecular Devices). Action potentials from the cells were recorded in current clamp mode using the same amplifier. Only cells with a stable resting membrane potential from the start of the patch were chosen for recording. Analysis was done using \textit{Axograph X} (Axograph Scientific). Peak currents and voltages were calculated by averaging 10 trials and Student's \textit{t}-test was used to determine statistical significance.\\

\section{Results}
\subsection{Signature Zernike modes in the neocortex and hippocampus}
Using fixed brain slice samples, we aimed to reduce the complexity for wavefront correction by identifying a few Zernike modes appropriate for optimizing the focus through different regions of the brain tissue as shown in Figure \ref{figure3}. We observe that in each of the 10 different positions in the sample, the hill climbing algorithm successfully produced an optimal focal spot after propagating through the sample. The aberration phase maps are made up of orthogonal Zernike modes without the need for tedious phase mapping (pixel by pixel) as demonstrated by \v{C}i\v{z}m\'ar, et al. \cite{cizmar2010situ}, which are unnecessary for individual brain slices.\\

\begin{figure}
\includegraphics[width=8.5cm]{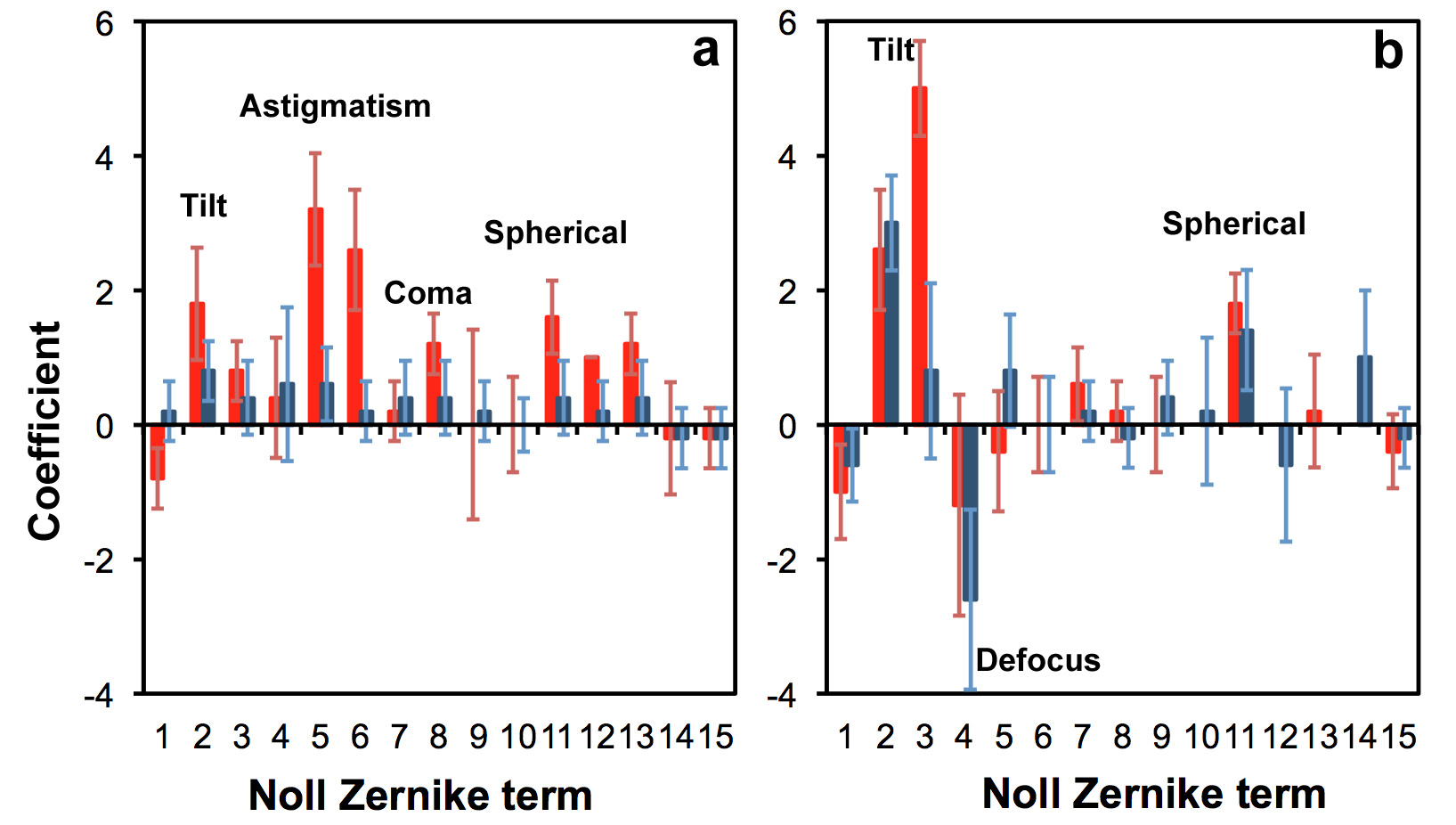}
 \caption{\label{figure4} Zernike modes for correcting light propagation through a (a) 100 $\upmu$m; and (b) 300 $\upmu$m thick parasagittal brain slice of rat.  The relevant regions were neocortex (red) and hippocampus (blue).   
 }
\end{figure}

Figure \ref{figure4} presents the corrections from each of 5 positions in the neocortex (red) and the hippocampus (blue) at two different tissue thicknesses of 100 $\upmu$m (figure \ref{figure4}a) and 300 $\upmu$m (figure \ref{figure4}b). For each trial (position in the sample), the 30 best corrections corresponding to the 30 highest intensity scores are obtained from the hill-climbing method.  Zernike modes are identified as `signature modes' if they consistently contribute to the best corrections. That is, signature modes are those that exhibit clear polarity with slight variances (e.g. NZT=5, coefficient= $3\pm1$).  Zernike modes that have large variances spanning to both positive and negative coefficients (e.g. NZT=10, coefficient= $0\pm1$) are instead interpreted as local correction modes that cannot be generalized within the studied region.\\

Figure \ref{figure4}a shows that at 100 $\upmu$m deep within the cortical region, astigmatism (NZT=5 and 6), coma-x (NZT=8) and spherical aberration (NZT=11) converge to consistent corrections, and are thus identified as signature modes (coefficients $\sim$3, 1 and 2, respectively). This is consistent with  Wang et al. \cite{wang2015direct} who recently reported the same modes (astigmatism, coma and spherical) dominating their corrective wavefront, obtained via guide star sensing. The tilt terms (NZT = 2 and 3) also converge to a positive coefficient, which is due to the surface flatness of the tissue as well a systematic tilt of the stage and are therefore not considered signature modes.  With thicker tissue (300 $\upmu$m), figure \ref{figure4}b shows that spherical correction (NZT=11) converges for both neocortex and the hippocampus, while tilt correction is more pronounced.  The other terms do not converge clearly compared to that of the 100 $\upmu$m thick sample.\\

\subsection{Efficient two-photon photolysis with wavefront correction}
From the previous experiment, we observed that each local correction relied heavily on just two or three out of the 12 available (non-tilt) Zernike modes, and we were able to identify signature modes corresponding to 100 $\upmu$m depth in the cortical region.  We then proceeded to use this prior information on wavefront correction for optimizing two-photon (2P) photolysis (uncaging) of caged glutamate at specific dendritic locations of neurons embedded within the brain slice. When released, the uncaged glutamate binds to receptors on the postsynaptic membrane causing a net flow of positive ions into the cell \cite{callaway1993photostimulation,denk1994two}. The resulting voltage response or excitatory postsynaptic potential (EPSP) can then be measured via the patch-clamp recording from the neuron.  \\

The non-linear absorption involved in 2P photolysis is similar to 2P fluorescence excitation where two low-energy (near-infrared) photons are required to release a caged glutamate. Similarly, a femtosecond (fs $\sim10^{-15}$s) pulsed laser focused by a high numerical aperture objective lens is necessary to provide a sufficiently high spatio-temporal photon density to bring about localized uncaging within a small focal volume. When traversing through a brain tissue, aberrations cause the focus to broaden and the efficiency of 2P photolysis is reduced. Efficacy of 2P photolysis can be observed by the changes in the EPSP where the maximum amplitude of the EPSP is associated to more caged neurotransmitters released as a result of a tighter focus.\\

\begin{figure}
\includegraphics[width=8.5cm]{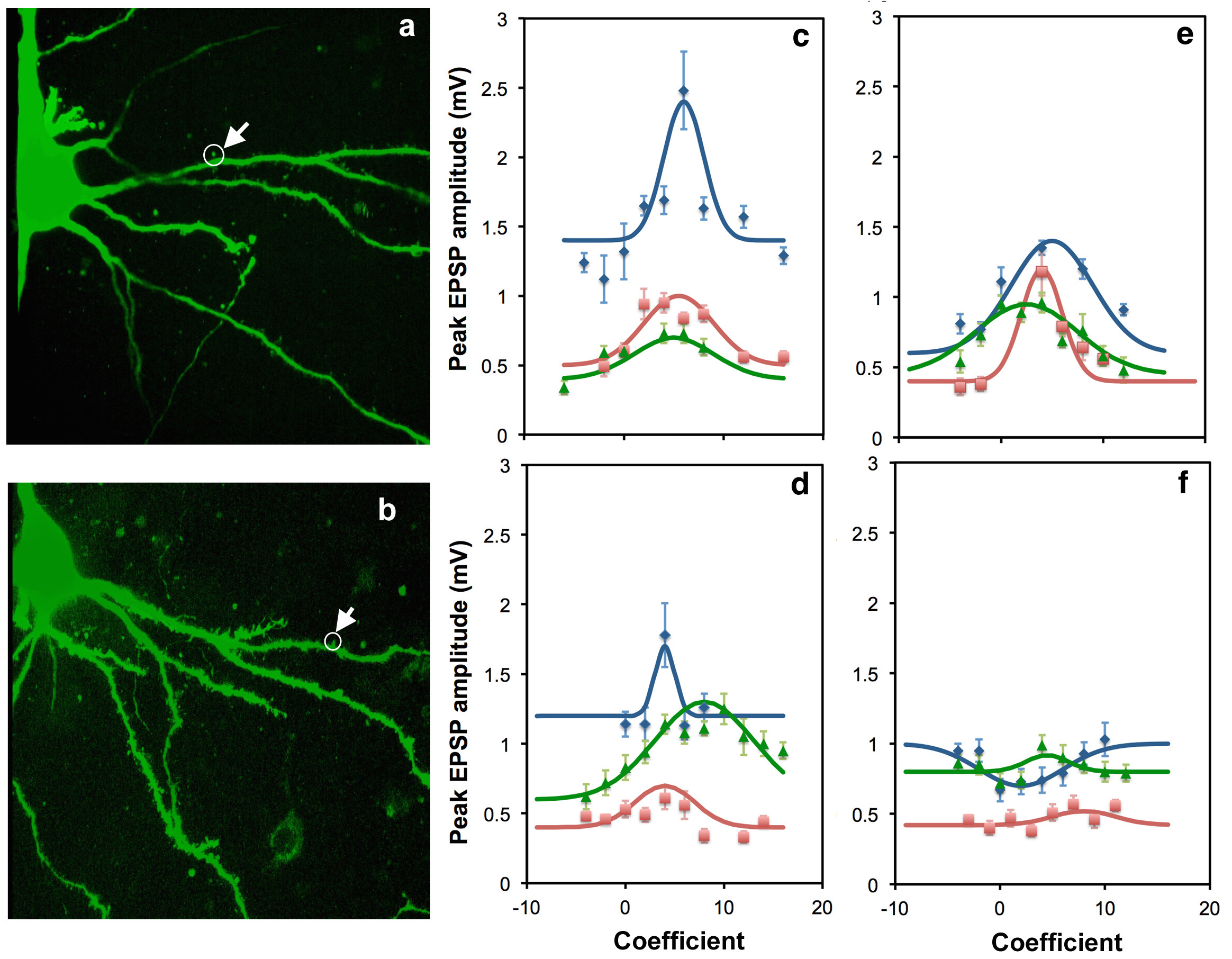}
 \caption{\label{figure5}Optimal two-photon (2P) photolysis of caged glutamate with wavefront correction. (a) and (b) Representative two-photon flattened stack images of neurons showing location of the dendritic spine targeted for 2P photolysis. Successful 2P photolysis optimization using Zernike modes for (c) astigmatism, (d) coma-x and (e) spherical aberration. (f) As expected, correction with coma-y did not yield consistent optimization for the different trials.}
\end{figure}

To optimize the efficiency of 2P uncaging within the brain tissue, the laser was encoded with signature Zernike modes in addition to lens and prism functions to position the focal spot in 3D \cite{go2013four}. Figure \ref{figure5} summarizes optimal uncaging following wavefront correction.  Figure \ref{figure5}a shows the two-photon image of the cortical pyramidal neuron (Layer V) showing the cell body on the left and basal dendrites extending towards the right.  The neuron's cell body is located around 50-70 $\upmu$m from the surface of the 300 $\upmu$m thick parasagittal slice.  The neuron's dendritic tree extends in three dimensions and may spread deeper into the tissue (e.g.  approx. 100 $\upmu$m deep).  We use the signature Zernike modes identified for a 100 $\upmu$m thick tissue (see figure \ref{figure4}a), namely astigmatism (NZT=5) (figure \ref{figure5}c), coma-x (NZT=8) (figure \ref{figure5}d) and spherical aberration (NZT=11) (figure \ref{figure5}e).  Unimodal optimization was observed for positive coefficient values in every animal trial,  as predicted. Each of the 12 curves correspond to a different Layer V pyramidal cell from a different animal, and each point is the average of 10 measurements of the peak EPSP. The errorbars represent the standard deviation. To show a negative result, we encoded the wavefront for correcting coma-y (NZT=7) and as expected no consistent optimization was observed for the three trials (figure \ref{figure5}f). \\

\section{Discussion}
Traditionally in astronomy, adaptive optics (AO) systems are employed to remove static aberrations in imperfect focusing elements and dynamic aberrations in atmosphere. These optical aberrations are decomposed into Zernike modes, which are part of an orthogonal basis set. In recent years, AO technologies have been gradually adopted by biomedical scientists to remove optical distortion caused by refractive index heterogeneity in biological tissue. However, many of the tissue-based AO optimization strategies borrow heavily from astronomy techniques. Astronomy imaging typically deals with fast-changing small refractive index fluctuation, but in tissue imaging, a much larger variation in the refractive heterogeneity with less rapid fluctuation is expected (not accounting for blood flow). Previously, Rueckel et al \cite{rueckel2006adaptive} and Zeng et al \cite{zeng20123d} began establishing tissue specific aberration optimization techniques based on fluorescence signals and Zernike modes. And more recently, Park et al \cite{park2015high} have demonstrated imaging brains through an intact skull. These techniques show the great potential in this field, however they all rely on complex measurement and correction of the scattered wavefront.\\

Using Zernike modes, we limit the optimization down to 15 modes, which is far more reduced compared to adjusting the magnitude of the phase (or coefficients) of an array of \textit{N} phase-shifting pixels (where \textit{N}$\sim$3228 ) \cite{vellekoop2007focusing}. We demonstrate that local wavefront corrections can be made using this finite set of Zernike modes. Furthermore, if we select specific regions within the brain slice, we further reduce the complexity to about 3 to 4 signature Zernike modes, effectively reducing an \textit{N} dimensional problem to 3 or 4 dimensions. Our experiment essentially shows that light scattering in brain tissues are not totally random but can be deterministic depending on the region of interest.\\

While our signature modes are specific to the brain regions and depths studied, the procedure can be extended to different areas of the brain and with other slice preparation (e.g. coronal slices as opposed to parasagittal).  Moreover, a similar procedure can also be used to map out aberrations for \textit{in vivo} observation of the brain. Hence, signature corrective Zernike modes for different types of experiments can be obtained to allow for direct optimization of the incoming wavefront, as an alternative to wavefront sensing and randomized wavefront correction optimization.  \\

\section{Conclusion}
The brain consists of a diverse set of neurons that are partially ordered forming  networks of neurons. These tissue structures, much like elsewhere in other organs, serve as a form of marker for identification under an optical microscope. Hence, light travelling through tissues (ballistic) is treated here as deterministic and therefore possesses a transmission matrix. While previous attempts have optimized light transmission through tissues, they have yet to investigate if a specific tissue structure would yield a signature transmission matrix. This study shows that even though local correction has a higher efficacy,  specific optical aberrations can be associated with specific regions of the brain tissue. The investigation of the Zernike mode optimization over spatial and axial positions provides a clear exposition of efficiency of the Zernike mode optimization in brain tissue slices and a simple methodology that can be adapted to most microscope systems with the addition of a wavefront shaping device on the excitation beam. \\

\appendix

\section{Calibration}

\begin{figure*}
\includegraphics[width=13cm]{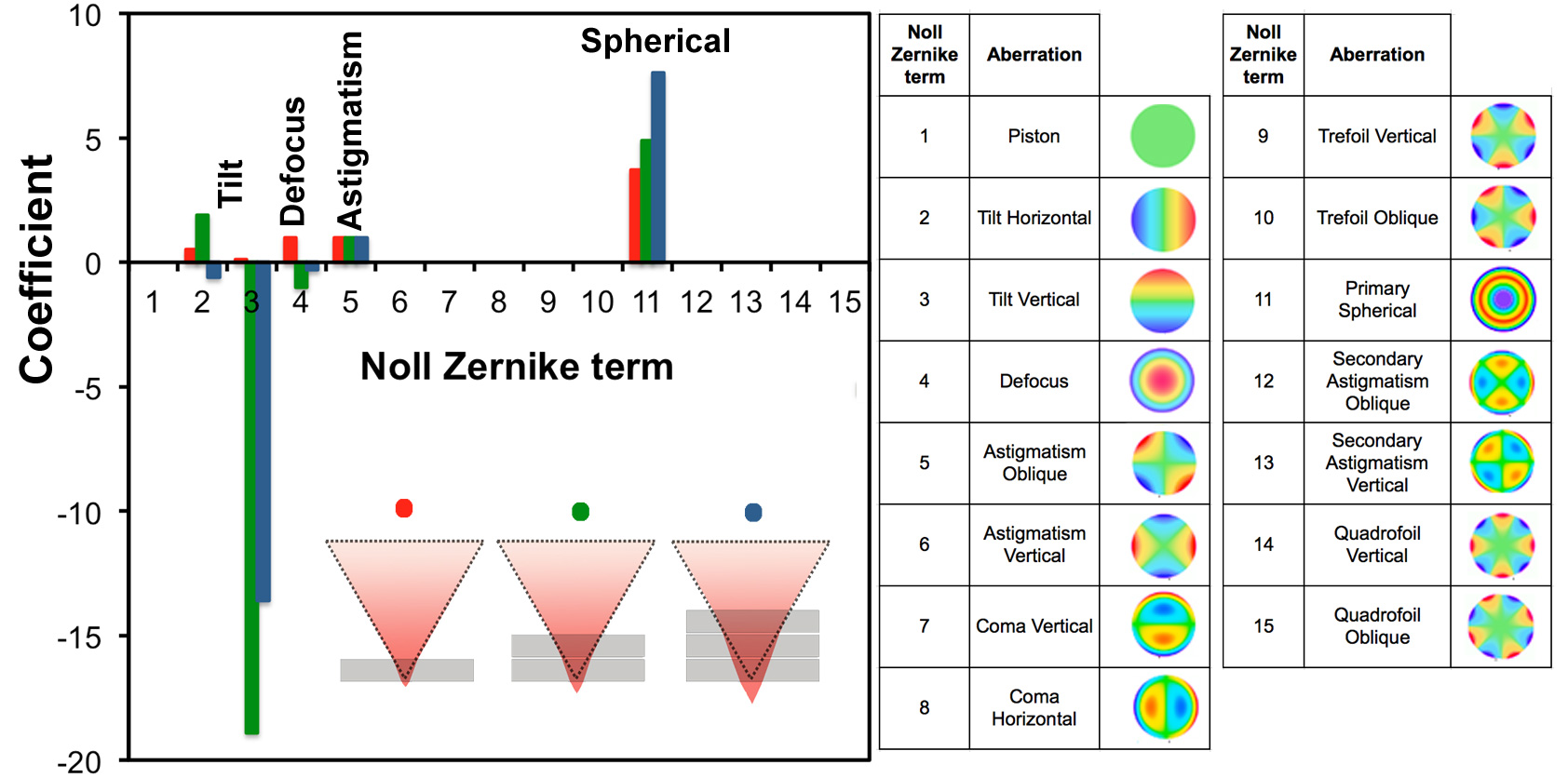}
 \caption{\label{figure6}Calibrating wavefront correction via Zernike modes with coverslips as samples.  (Left) The coverslips introduce known aberrations (e.g. spherical aberration) as depicted in the plot with a positive coefficient (approx. 5-8) for Noll Zernike term NZT=11 which is required to correct for spherical aberration.  (Right) Zernike modes from NZT=1 to 15 (coefficient=1) with corresponding aberration and phase pattern.
 }
\end{figure*}
\begin{figure*}
\includegraphics[width=13cm]{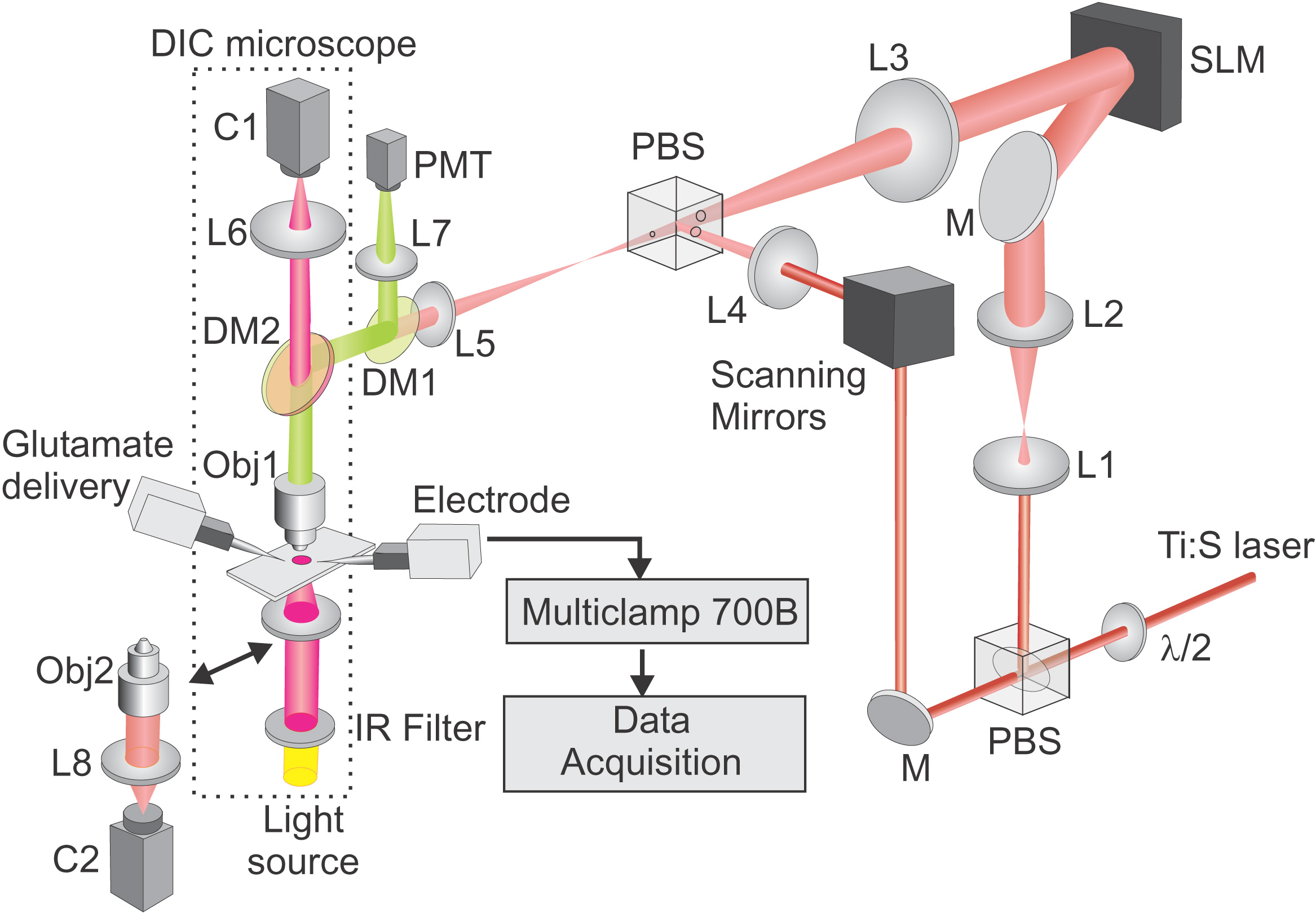}
 \caption{\label{figure7}Optical setup showing two Ti:S laser illumination modes: (1) for a two-photon scanning microscope via the scanning mirrors; and (2) for a wavefront encoded beam via the spatial light modulator (SLM).  Both illumination modes are combined for focal illumination to the sample via the objective lens (Obj1).  The microscope is built from a conventional differential interference microscope (Olympus BX50) using camera, C1.  For \textit{a priori} identification of signature Zernike modes, the focus is imaged by Obj2 and onto camera, C2. Other components: mirror (M); lens (L); half-wave plate ($\lambda$/2); dichroic mirror (DM); photomultiplier tube (PMT); and polarizing beam splitter (PBS).  }
 \end{figure*}

To calibrate the system, we first optimized the focus through optical materials exhibiting a known aberration. A stack of microscope coverslips is known to introduce spherical aberration \cite{ji2010adaptive}. We increased the aberrations by stacking microscope coverslips of equal thickness and refractive index without altering the position of the two objectives. This results in the paraxial rays and the marginal rays having a larger path length difference that lead to larger spherical aberrations. In figure \ref{figure6} (left), we show the imposed wavefront decomposed into its respective Zernike modes for optimizing through 1 to 3 glass coverslips. The coefficients on the y-axis describe the magnitude of wavefront correction using a particular Zernike mode. The Zernike modes that showed a clear improvement in the focus are spherical aberration (NZT=11) and defocus (NZT=4). Figure \ref{figure6} (left) also shows a systematic astigmatism correction (NZT=5) caused by slight misalignment in the optical system.  Note that the coefficient for astigmatism does not change with increasing number of coverslips. Corrections to tilt (NZT=3 and 4) are caused by the angle of the stage with respect to the optical axis. When stacking the coverslips, slight variations in coverslip flatness cause some amount of tilt uncertainty.  The rest of the Zernike modes do not contribute to the optimization of the focus.  Figure \ref{figure6} (right) shows the phase patterns of the various Zernike modes (coefficient=1) used in this work. 

\section{Slice Preparation}

We prepared parasagittal brain slices from 15 to 19 day old Wistar rats. We followed the standard procedures for animal experiments in accordance with the methods approved by the Animal Experimentation Ethics Committee of the Australian National University, Australia. The slices were cut with a vibratome (Leica VT1200S) in ice-cold oxygenated artificial cerebrospinal fluid (ACSF) that contained (in mM): 1.25 NaH$_2$PO$_4$, 1.0 MgCl$_2$, 125.0 NaCl, 2.5 KCl, 2.0 CaCl$_2$, 25.0 NaHCO$_3$, and 10.0 glucose. Slices were incubated in oxygenated ACSF at 34$^{\circ}$C for 30 min and kept at room temperature before being fixed or transferred to the recording chamber. For \textit{a priori} identification of signature Zernike modes, we cut 100 $\upmu$m and 300 $\upmu$m thick brain slices, which were fixed in 4\% paraformaldehyde, embedded in Mowiol 4-88 medium and placed in between two type-0 coverslips.\\

To demonstrate effective photostimulation of neurons, we prepared 300 $\upmu$m thick brain slices and targeted pyramidal cells in layer V in the somatosensory cortex. The neurons were patched with a glass electrode (resistance: 4-6 M$\Omega$) containing (in mM): 115 K-gluconate, 20 KCl, 10 HEPES, 10 phosphocreatine, 4ATP-Mg, 0.3GTP, 5.4 biocytin, and 0.3 AlexaFluor-488 (Sigma-Aldrich).  Recording of the neuron's postsynaptic potentials was performed using a MultiClamp 700B amplifier (Molecular Devices) and analyzed using \textit{Axograph X} (Axograph Scientific). \\

\section{Optical Setup} 
 
Figure \ref{figure7} shows the multi-functional two-photon microscope setup \cite{go2013four} where a femtosecond pulse Ti:S laser (Coherent Inc. MIRA 900 pumped with 12 W Coherent Verdi G12) is split via a polarizing beam splitter (PBS1).  The S-polarized beam is directed to a xy-galvano-scanning mirror setup and used for two-photon imaging. PBS2 couples the S-polarized beam to form a two-photon microscope together with lenses (L4, L5 and L7), dichroic mirrors (DM1) and the photomultiplier tube (PMT) to detect the fluorescence from the sample. On the other hand, the P-polarized laser beam is directed towards an SLM via a series of mirrors and beam expansion lenses, L1 and L2. A linear combination of Zernike modes is generated by the computer algorithm and displayed on the SLM, which introduces a relative phase delay on the incident light proportional to the brightness of the pixel. The encoded phase map is relayed via a 4f-lens configuration (L3 and L4) to the back aperture of the objective lens (Obj1). The appropriate wavefront correction at the back aperture of the objective lens projects a corrected focus at the Fourier plane. The P-polarized beam is also coupled to the microscope using PBS2. For \textit{a priori} identification of Zernike modes, the laser spot focused at the bottom of the brain slice is imaged onto a CCD camera and digitally recorded after passing through another objective lens (Obj2), L8 and onto a camera, C2.  During electrophysiological experiments with living neurons, the setup at the bottom is replaced with polarized illumination for differential interference microscopy (DIC) together with L6 and C1. Using a DIC microscope allows for easier patching of the glass electrode onto the neuron's membrane.

\subsection{Acknowledgments}
This project has been funded under the Australian Research Council Discovery Project (ARCDP 140101555).\\\\

\bibliography{ref}

\end{document}